\documentclass[preprint, 12pt, 3p,times]{elsarticle}
\usepackage{graphicx}
\usepackage{amsmath}
\usepackage{amsfonts}
\usepackage{mathrsfs}
\usepackage[OT1]{fontenc} 
\usepackage{enumerate}
\usepackage[inline]{enumitem}
\usepackage{amsthm}
\usepackage{tabularx}
\usepackage{arydshln}

\usepackage[colorlinks=true,linkcolor=blue,citecolor=blue]{hyperref}
\usepackage{mwe}
\biboptions{authoryear, sort}
\usepackage{subfig}
\usepackage{float}
\usepackage{graphicx}
\usepackage{textcomp}
\usepackage{footnote}
\usepackage{makecell}
\usepackage[dvipsnames]{xcolor}
\usepackage[draft]{todonotes}
\usepackage{slashbox}

\colorlet{revised}{Red}

\begin{document} 

\begin{frontmatter}

\title{The Effectiveness of Managed Lane Strategies for the Near-term Deployment of Cooperative Adaptive Cruise Control}
%\tnotetext[label0]{This is only an example}

\author[label1]{Zijia Zhong\corref{cor1}}
\address[label1]{Department of Civil and Environmental Engineering, University of Delaware}
%\address[]{}
%\address[]{Newark, DE}

\cortext[cor1]{Corresponding author: Zijia Zhong}
%\fntext[label3]{I also want to inform about\ldots}
%\fntext[label4]{Small city}

\ead{zzhong@udel.edu}
%\ead[url]{author-one-homepage.com}

\author[label2]{Joyoung Lee}
\address[label2]{John A. Reif, Jr. Department of Civil and Environmental Engineering, New Jersey Institute of Technology}
\ead{ jo.y.lee@njit.edu}

\begin{abstract}
Traffic simulation is a cost-effective way to test the deployment of Cooperative Adaptive Cruise Control (CACC) vehicles in a large-scale transportation network.  By using  a previously developed microscopic simulation testbed, this paper examines the impacts of four managed lane strategies for the near-term deployment of CACC vehicles under mixed traffic conditions.  Network-wide performance measures are investigated from the perspectives of mobility, safety, equity, and environmental impacts. In addition, the platoon formation performance of CACC vehicles is evaluated with platoon-orientated measures, such as the percentage of platooned CACC vehicles, average platoon depth, and vehicle-hour-platooned that is proposed in this paper under the imperfect DSRC communication environment. Moreover, managed lane score matrices are developed to incorporate heterogeneous categories of performance measures, aiming to provide a more comprehensive picture for stakeholders. The results show that mixing CACC traffic along with non-CACC traffic across all travel lanes is an acceptable option when the market penetration (MP) is lower than 30\% for roadways where a managed lane is absent. Providing CACC with priority access to an existing managed lane, if available, is also a good strategy for improving the overall traffic performance when the MP is lower than 40\%. When the MP reaches above 40\%, a dedicated lane for CACC vehicles is recommended, as it provides greater opportunity for CACC vehicles to form platoons. The facilitation of homogeneous CACC traffic flow could make further improvements possible in the future.
\end{abstract}

\begin{keyword}
%% keywords here, in the form: keyword \sep keyword
Managed Lane\sep Cooperative Adaptive Cruise Control\sep Microscopic Traffic Simulation\sep Platoon Formation \sep Mixed Traffic Conditions\sep Imperfect DSRC Communication
%% MSC codes here, in the form: \MSC code \sep code
%% or \MSC[2008] code \sep code (2000 is the default)
\end{keyword}

\end{frontmatter}

%%
%% Start line numbering here if you want
%%
% \linenumbers

%% main text
\section{Introduction}
\label{sec1}
Connected and Automated Vehicle (CAV) technology will revolutionize the way vehicles are operated.  Among its wide-ranging applications, Cooperative Adaptive Cruise Control (CACC) is identified as one of the thrust applications by governments, industries, and academia all around the world in improving mobility, environment, and more importantly safety. The field experiments of CACC systems have been accelerated as the technology matures in recent years.  The highlights of the deployments, to name a few, include the United State Department of Transportation (USDOT)-approved CAV test sites \citep{xiang2009wireless}, the California PATH's Automated Highway System (AHS) program  \citep{rajamani2001experimental,bu2010design,milanes2014cooperative}, the Energy ITS Project \citep{ITSEnergy2011}, SARTRE    \citep{van2013multi}, KONVOI \citep{wille2007konvoi}, and the European Truck Challenging 2016 \citep{eckhardt2016european}. Despite policy encouragement from government agencies and the progress made in academic research, large-scale field deployment, however, are still considered premature at the current stage, in terms of safety, technological, and budgetary concerns. Simulation is one of the best approaches to bridge the gap between prototype testing CACC technology and the large-scale real-world deployment by providing a virtual environment. Simulation can also improve the quality of the testing by expanding the testing scope to a wide spectrum of scenarios without incurring a significant amount of additional cost. 

Managed lanes (MLs), as defined by the Federal Highway Administration (FHWA), are highway lanes that are set aside and operated under a variety of fixed and/or real-time strategies in response to local goals and objectives \citep{kim2015sustainable}, such as traffic operation improvement, emission reduction, safety enhancement etc. As solutions for efficient lane management, ML strategies have been implemented nationwide with various forms (e.g., high-occupancy vehicle (HOV) lane, high-occupancy toll (HOT) lane, exclusive bus lane (XBL)). Pairing ML strategies with CACC (designated as CACC-ML) becomes a logical step in promoting CACC, because of CACC's distinct operational characteristics in comparison to human-driven vehicles (HVs). Furthermore, the provision of CACC-ML can provide road users with stronger incentives either to purchase CACC vehicles that meet the minimum functional requirements or to factor in the option of retrofitting certified aftermarket CACC systems to a reasonable extent. 

In this paper, the effectiveness of deploying CACC-ML strategies is examined with the goal of deriving policy recommendations for the initial introduction of CACC vehicles into mixed traffic.  The organization of the remainder of the paper is as follows.  Relevant research regarding deploying CACC vehicles in mixed traffic is reviewed in Section \ref{Literature Review}, followed by the microscopic simulation framework in Section \ref{Methodology}. The simulation results are presented and discussed in Section \ref{Simulation Result}. Conclusions and future work are discussed in Section \ref{Conclusions}.

\section{Literature Review}
\label{Literature Review}
Increasing practices of MLs have been observed over the years and case studies show that the desired travel patterns or behaviors can be incentivized by ML strategies. For instance, the California Clean Air Vehicle Decal program \citep{shewmake2014hybrid} was designed to promote user adoption of energy-efficient and low-emission hybrid vehicles by allowing the owners to drive on HOV lanes without even meeting the occupancy requirement. It is believed that the adoption of CACC vehicles can be facilitated with policy alike. The CACC-ML provision is envisioned to have three phases. First, the adoption of CACC vehicles is incentivized by allowing the use of the ML free of charge. At this stage, the following headway of CACC vehicles on the ML may be comparable to that has been observed from HVs for safety reason in mixed traffic. Then a shorter following headway for CACC vehicles could be implemented to further increase the carrying capacity of the ML when the demand as well as the familiarity of road users to CACC increases.  Lastly, when the CACC vehicles reach a critical level of MP, the transition to CACC-only lane could be eventually made. At this stage, high-performance driving enabled by CACC can be achieved because of a homogeneous CACC traffic on the ML \citep{ZhongDis}.

The benefits of CACC for enhancing the performance of a freeway system have been reported. The capacity of 4,250 vph per lane was observed by \cite{van2006impact} on a 6-km segment with uniformly distributed ramps under full market penetration. \cite{shladover2012impacts} concluded the capacity of 3,600 vph per lane at 90\% MP in the absence of on-ramp demand for a one-lane freeway. \cite{Arnaout2014} evaluated CACC under moderate, saturated, and over-saturated demand on a hypothetical 4-lane highway under different market penetrations. They found that 9,400 vehicles (out of 10,000 vehicles) could be served in an hour when the CACC reached 40\% MP. \cite{lee2014mobility} found the mobility benefits of CACC began to show at 30\% MP. \cite{songchitruksa2016} assessed the improvements of CACC on the 26-mile I-30 freeway in Dallas, TX and observed the highest throughput being 4,400 vph at 50\% MP.

Insofar, studies on MLs for CAVs are relatively scarce in the literature. Traditional lane management experience may not be directly transferrable to CACC deployment because of the the difference in the underlying operational principles.  
%The communication capability of CAV could in turn promote the effectiveness of the ML by furnishing supportive real-time information to travelers and to roadway authroities for real-time management.    
There are two major approaches for studying CACC-ML: analytical modeling and computer simulation. \cite{hussain2016freeway} proposed an analytical ML model for determining the optimal lanes to be allocated for CAV traffic under single-lane and ML environment. Three types of CAV following headway (conservative, neutral, and aggressive) were assigned. The maximization of the overall system throughput could be computed under the assumption that both CAVs and HVs are randomly distributed in a freeway facility. From a macroscopic perspective, \cite{qom2016evaluation} used simulation-based dynamic traffic assignment to study the diversion of CACC vehicles onto a ML with toll incentives from dynamic pricing. Using Vissim, \cite{zhang2018operational} investigated the operational capacity of CACC-ML with dedicated CACC on/off ramps. It was discovered that the ramp density should be placed no less than 5 km (3 miles) apart to ensure a reasonable level of throughput. 
\cite{Liu2019Modeling} found that deploying the ML and vehicle awareness device (VAD)-equipped vehicles were proven to be helpful under low and medium  market penetrations. The increase in the pipeline capacity (under the ideal condition with  no lane changes, uniform desired speed, etc.) was estimated between 8\% to 23\%. On a subsequent test on an 18-km segment of the SR-99, they observed great improvement on traffic flow pattern when the MP was at 40\% or higher on the corridor-level speed heat map.

%\textcolor{Gray}{
A CACC vehicle is able to operate in a safe manner with a much smaller following headway due to the V2V communication with its predecessor(s) \citep{shladover2018using,tsugawa2016review, Nowakowski2011}. Clustering CACC vehicles, therefore, becomes a crucial task in operation.   Note that CACC ``clustering'' and ``platooning'' are used interchangeably.  According to \cite{shladover2015cooperative}, three types of clustering strategies   could be expected in operation: 1) ad hoc clustering, 2) local coordination, and 3) global coordination.  Ad hoc clustering assumes that the CACC vehicles arrive in random sequence and the CACC vehicles do not actively seek clustering opportunities. Therefore, the probability of driving around other CACC vehicles is highly correlated to MP. The ad hoc clustering has been adopted in most of the previous studies thus far because of its simplicity on implementation. 
%}

%\begin{table}[]
%\centering
%\caption{Headway for intra-platoon following} 
%%\resizebox{\textwidth}{!}
%%{
%\begin{tabular}{p{1in}|p{1in}|p{4in}} \hline 
%Intra-platoon HW & Inter-platoon HW & Study\\ \hline 
%0.5  & & \cite{vander2002effects, van2006impact, Arnaout2014}\\
%0.6 &  & \cite{songchitruksa2016,shladover2012impacts,NAP25366, lee2014mobility} \\
%0.7 & & \cite{songchitruksa2016, shladover2012impacts, Liu2019Modeling} \\
%0.8 & & \cite{NAP25366}\\
%0.9 & & \cite{songchitruksa2016,shladover2012impacts}\\
%1.0 & & \cite{NAP25366}\\
%1.1 & & \cite{songchitruksa2016,shladover2012impacts}\\
%1.2 & & \cite{lee2014mobility} \\
%1.4 & & \cite{vander2002effects}\\
%\hline
%\end{tabular}
%%}
%\label{table:vissimParameters}
%\end{table}
	
Local coordination, commonly known as active platooning, could be employed to help with clustering. Under this concept, CACC vehicles actively identify and approach an existing CACC cluster or other free-agent CACC vehicles to form a new cluster. The local coordination has been discussed by \cite{liang2013fuel}, \cite{davila2012sartre}, and \cite{lee2014mobility}.  One of the main challenges to local coordination is the determination of the relative position of a vehicle with sufficient accuracy and reliability. Solutions have been proposed to aid the localization of CACC vehicles \citep{shladover2015cooperative}, for instance, 1) infrastructure lane identification by radio frequency identification (RFID), 2) vehicle-based lane identification by on-broad sensors (e.g., GPS, inertial measurement unit, and camera), 3) vehicle-based confirmation of the predecessor by using visual or infrared-camera-visible marking, and 4) visual confirmation by the drivers.  Lastly, global coordination uses advance planning to coordinate vehicles traveling with the same origin-destination pair even before the CACC vehicles entering the highway \citep{larson2013coordinated}. The most likely application for global coordination would be long-haul trucking or lengthy commute trip on congested highways. 
	
Equity issues is another crucial aspect for MLs, and it has to be properly addressed, due to its vital role in gaining public support. The perception of exclusivity for different road users could be a sensitive matter. \cite{vtpi2018} conducted extensive research regarding the equity and social justice for the impact of a ML in CAV context. It identified that a dedicated lane for CACC could impact various social classes \citep{litman2017autonomous}, since not too many people are able to afford CACC vehicles.  The issue becomes more pronounced when speed limit privilege (e.g., higher speed limit) is assigned to CACC lane. Hence, the impacts of a CACC lane among users and non-users need to be quantified for further assessment. More mobility options could be provided (e.g., access to transit) in order to gain the benefits of CACC-ML for different user groups. Learned from existing MLs, additional considerations include: 1) cost incentives to make CACC accessible for all via tax deduction, subsidies, etc.; 2) possible shared mobility for CACC vehicles; and 3) availability of alternative routes for other road users.

The CACC-ML strategy could provide technical accommodation, economic incentive, and mobility improvement for users who decide whether to purchase CACC-equipped vehicles. CACC-ML could be instrumental in the near-term deployment of CACC. The effectiveness of CACC-ML strategies can be evaluated systematically in a cost-efficient way with microscopic simulation.

\section{Methodology}
\label{Methodology}
%Simulation is also able to provide flexibility for exploring ML policies that might not be acceptable due to potential ramifications.
\subsection{Simulation Framework}
\begin{figure} [ht!]
	\centering
	\includegraphics[width=0.75\textwidth]{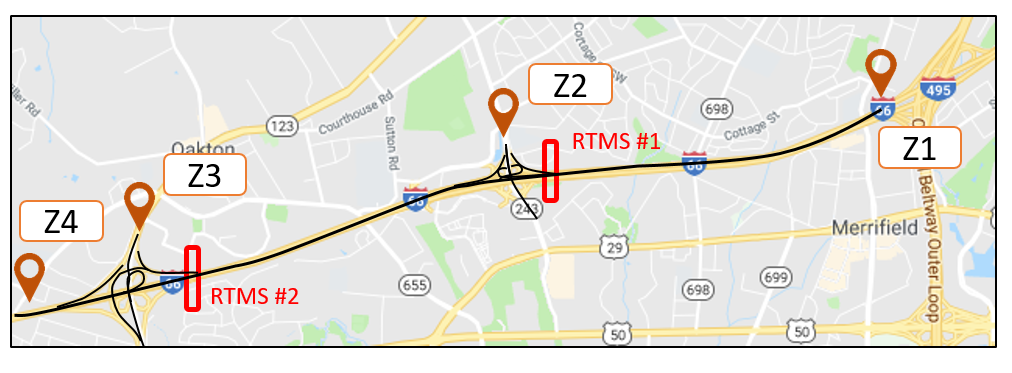}   
	\caption{Interstate highway 66 (I-66) segment}
	\label{fig:i66}
\end{figure}

A segment of a major commuter corridor, the Interstate Highway 66 (I-66) located in Fairfax County, Virginia, is used to evaluate the potential operation impacts for deploying CACC vehicles in the near term.  The 8-km (5-mile) stretch of the roadway (shown in Fig. \ref{fig:i66}) is located outside of the Washington D.C. Beltway (I-495) with four lanes in each direction and two interchanges. Currently, an HOV lane is implemented in the leftmost lane in each direction. No physical barrier is in place between the General Purpose Lanes (GPLs) and the HOV lane. 
The I-66 network has been used in previous studies by  \cite{Li2019High}, \cite{NAP25366}, \cite{ZhongDis}, and \cite{Zhong2018}. The calibration was conducted with three independent data sources:  INRIX travel time, remote traffic microwave sensor data (RTMS) flow data, and video camera data to ensure a realistic representation of the real-world traffic conditions \citep{STOLT4, Li2019High}. The first two were the primary sources of fine-tuning the simulation network. An overview of the traffic condition for each origin is shown in Table \ref{table:trafficCondition}. The demand is defined in a 15-min interval. 

% \citep{Li2019High, NAP25366, ZhongDis, Zhong2018}
\begin{table}[!ht]
\centering
\caption{Independent Data Source for Calibration}
\begin{tabular}{cccccc}
\hline  
Data Source &  Volume & Speed & Occupancy & Travel Time & Measurement Type \\ \hline
RTMS data & \checkmark & \checkmark & \checkmark & & point\\
INRIX data & & & & \checkmark & segment\\
Video camera footage & \checkmark  & & & & point\\
\hline 
\end{tabular}
\label{table:dataSource}
\end{table}

\begin{table}[!ht]
\centering
\caption{Network Traffic Demand}
\begin{tabular}{cccc}
\hline 
Zone ID &  Description & Peak/Avg. GP Demand, vph & Peak/Avg. HOV Demand, vph \\ \hline
Z1 & I-66 East or I-495 & 4197/3998 & 1885/1796\\
Z2 & Exit 62 Nutley St. & 712/382 & 335/180\\
Z3 & Exit 60 SR 123 & 1411/1146 & 1411/1146\\
Z4 & I-66 West & - & -\\
\hline 
\end{tabular}
\label{table:trafficCondition}
\end{table}

\begin{figure} [ht!]
	\centering
	\includegraphics[width=\textwidth]{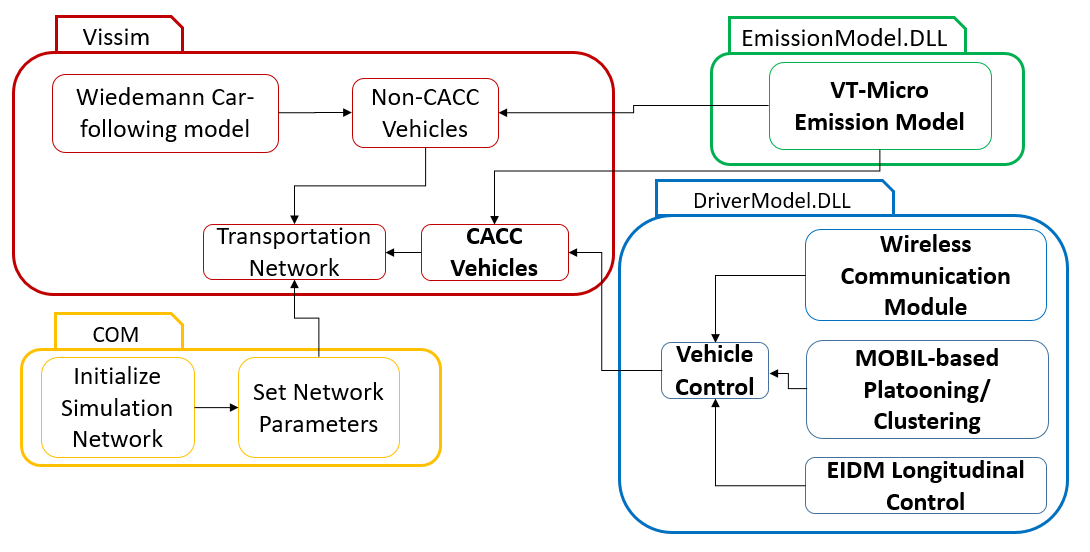}   
	\caption{Vissim-based microscopic simulation testbed}
	\label{fig:simFrame}
\end{figure}

The simulation framework is developed based on PTV Vissim \citep{ptv2018}, Vissim Component Object Model (COM) interface \citep{vissim2015introduction}, and the Vissim DriverModel.DLL application programming interface \citep{VissimDllDoc} (VEDM thereafter). The overall simulation framework is shown in Fig. \ref{fig:simFrame}. Vissim is a multimodal microscopic traffic simulation software, where each entity (e.g., car, train, or pedestrian) is simulated individually. Such capability is one of the most crucial elements for simulation of CACC vehicles.  The VEDM is programmed to replace the default Wiedemann car-following model \citep{wiedemann1991} with the custom CACC behavioral model. 

\begin{figure} [H]
	\centering
	\includegraphics[width=\textwidth]{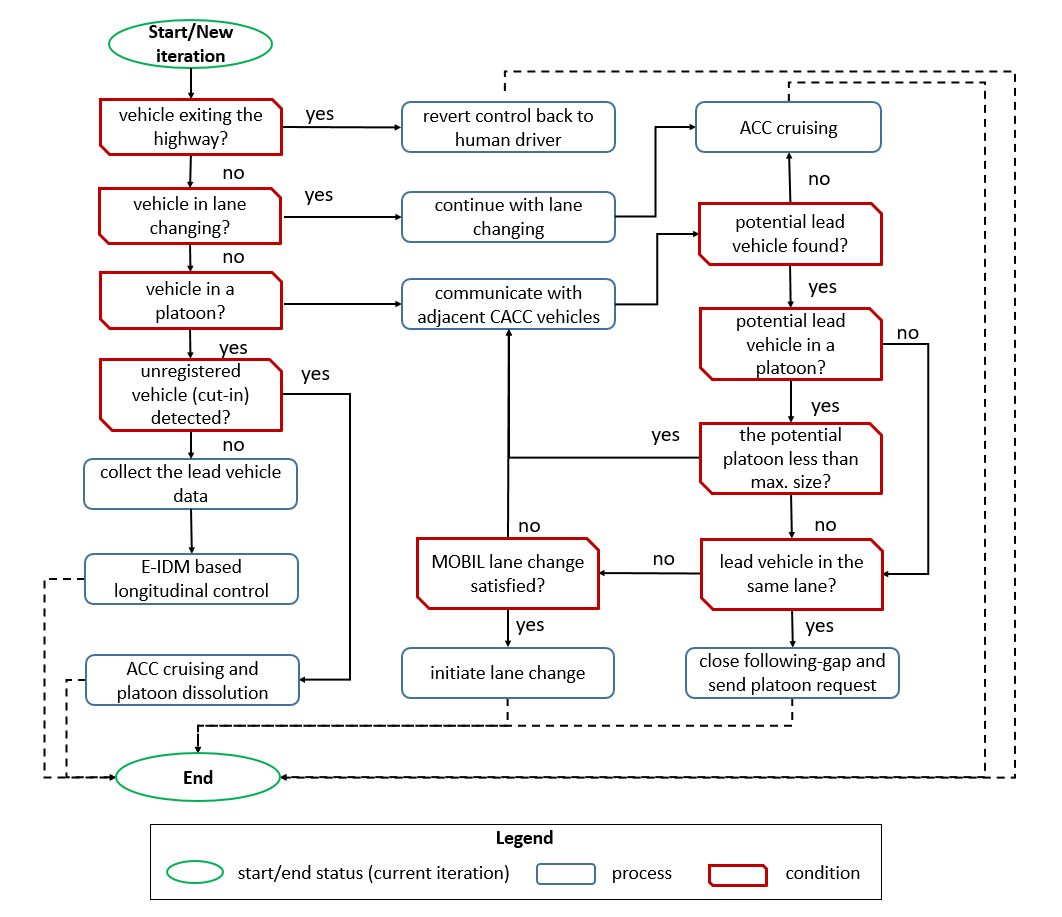}   
	\caption{\textcolor{revised}{CACC clustering algorithm}}
	\label{fig:pltnForm}
\end{figure}

The CACC vehicles are equipped with automated longitudinal control that is based on the Enhanced Intelligent Driver Model (E-IDM) \citep{kesting2010enhanced}. The behavior model of the EIDM is expressed in Eq. \ref{eq:eidm}, \ref{eq: minDistCal}, and \ref{eq: cahCal}.
\begin{equation}
\ddot{x}=\begin{cases}
a[1-(\frac{\dot{x}}{\dot{x_{des}}})^{\delta }- (\frac{s^{*}(\dot{x}, \dot{x}_{lead})}{s_{0}})] & \text{ if } x=  \ddot{x}_{IDM} \geq \ddot{x}_{CAH} \\ 
 (1-c)\ddot{x}_{IDM} + c[\ddot{x}_{CAH} + b \cdot tanh ( \frac{\ddot{x}_{IDM} - \ddot{x}_{CAH}}{b})] & \text{otherwise} 
\end{cases}
\label{eq:eidm}
\end{equation}
\begin{equation}
s^{*}(\dot{x}, \dot{x}_{lead}) = s_{0} + \dot{x}T + \frac{\dot{x}(\dot{x} - \dot{x}_{lead})}{2\sqrt{ab}} 
\label{eq: minDistCal}
\end{equation}
\begin{equation}
\ddot{x}_{CAH}=
\begin{cases}
\frac{\dot{x}^{2} \cdot \min(\ddot{x}_{lead}, \ddot{x})}{\dot{x}_{lead}^{2}-2x \cdot \min(\ddot{x}_{lead}, \ddot{x})} & 
\dot{x}_{lead} (\dot{x} - \dot{x}_{lead}) \leq -2x \min(\ddot{x}_{lead}, \ddot{x})  \\
\min(\ddot{x}_{lead}, \ddot{x}) - \frac{(\dot{x}-\dot{x}_{lead})^{2} \Theta (\dot{x}- \dot{x}_{lead})}{2x}  & \text {otherwise}
\end{cases} 
\label{eq: cahCal}
\end{equation}
where, $a$ is the maximum acceleration; $b$ is the desired deceleration; $c$ is the coolness factor; $\delta$ is the free acceleration exponent; $\dot{x}$ is the current speed of the subject vehicle;  $\dot{x}_{des}$ is the desired speed,  $\dot{x}_{lead}$ is the speed of the lead vehicle; $s_{0}$ is the minimal distance; $\ddot{x}$ is the acceleration of the subject vehicle; $\ddot{x}_{lead}$ is the acceleration of the lead vehicle; $\ddot{x}_{IDM}$ is the acceleration calculated by the original IDM model \citep{Treiber2000}. The minimal distance can be calculated by Eq.\ref{eq: minDistCal} where  $T$ is the desired time gap; and $\ddot{x}_{CAH}$ is the acceleration calculated by the constant-acceleration heuristic (CAH) component as shown in Eq.\ref{eq: cahCal}, where $\Theta$ is the Heaviside step function. The local coordination of CACC vehicles is incorporated into the VEDM as well, where a free-agent CACC vehicle (that is not in a platoon) detects its surrounding traffic, trying to join an existing platoon or form a platoon with other free-agent CACC vehicles. Thus far, three types of platoon formation have been used: front-, mid-, and rear- joining \citep{lee2014mobility, Amoozadeh2015}. In this study, only rear-joint is allowed for jointing a platoon in order to control the variables of the evaluation.  The decision-making procedure for a free agent CACC vehicle is exhibited in Fig. \ref{fig:pltnForm}. 

%%-------------- the addtion MOBIL description is below. Save for later.
The Minimizing Overall Braking Induced by Lane Change model (MOBIL) \citep{kesting2007mobil} as expressed in Eq.\ref{eq: mobil} is adopted as the mechanism to prevent lane changing of a free-agent CACC vehicle that is potentially disruptive to the surrounding traffic. 
\begin{equation}
\tilde{\ddot{x}} - \ddot{x} + p(\tilde{\ddot{x}}_{n}-\ddot{x}_{n} + \tilde{\ddot{x}}_{o} - \ddot{x}_{o}) > \Delta \ddot{x}_{th} 
\label{eq: mobil}
\end{equation}
where $p$ is the politeness factor; $\tilde{\ddot{x}}$ is the new estimated acceleration for the subject vehicle; $\ddot{x}$ is the current acceleration of the subject vehicle; 
$\tilde{\ddot{x}}_{n}$ is estimated acceleration of the new trailing vehicle; $\ddot{x}_{n}$ is the previous acceleration of the new trailing vehicle; $\tilde{\ddot{x}}_{o}$ is the new estimated acceleration for the old trailing vehicle; $\ddot{x}_{o}$ is the previous acceleration of the old trailing vehicle. Note that the estimated acceleration is calculated by the E-IDM model.

\begin{figure} [h]
	\centering
	\includegraphics[width=0.9\textwidth]{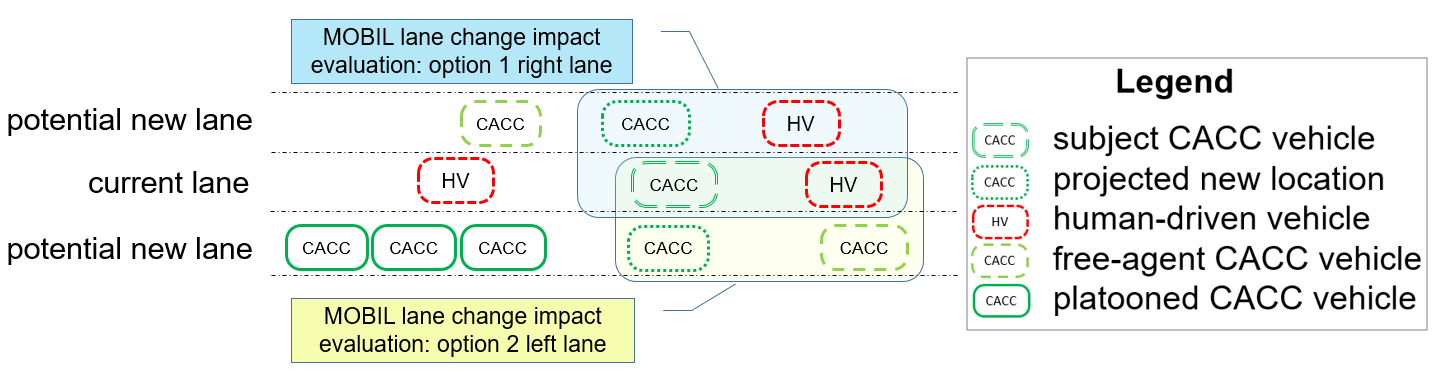}   
	\caption{MOBIL-based lane change decision}
	\label{fig:mobileDemo}
\end{figure}

When a potential platooning opportunity is identified via V2V communication, the CACC system estimates the impacts on the immediate vehicles based on MOBIL should the lane change be initiated.  A lane change is only executed when the expected negative impact is below the pre-determined threshold, $\Delta \ddot{x}_{th}$. In the event of multiple clustering alternatives,  the one with the smallest value (less impact) in the left hand side of the Eq.\ref{eq: mobil} is chosen, as illustrated in Fig. \ref{fig:mobileDemo}.

The parameters for the E-IDM  models are shown in table \ref{table: parameters}. Note that two types of headway are implemented in the longitudinal control. The $T_{intra}$ represents the intra-platoon headway, which is typically smaller owing to the communication within a CACC platoon. The larger headway, $T_{inter}$, is used when a CACC vehicle follows a non-CACC vehicle. The minimum number of CACC vehicles for a platoon is set as three and the maximum number of CACC vehicles for a platoon is set as seven. 
\begin{table}[!ht]
\centering
\caption{CACC Vehicle Control Parameters} 
\begin{tabular}{ccccccccccccc}
\hline 
Parameter & $T_{intra}$ & $T_{inter}$ & $s_{0}$ & $a$ & $b$ & $c$ & $\theta$ & $\dot{x}_{des}$ & $p$ & $\Delta \ddot{x}_{th}$ \\ \hline
value & 1 s & 1.2 s & 1$m$ & 2$m/s^{2}$ & 2$m/s^{2}$ & 0.99 & 4 & 105 $km/h$ & 0.9 & 1$m/s^{2}$ \\
\hline 
\end{tabular}
\label{table: parameters}
\end{table}

\begin{figure}[H]
	\centering
	\includegraphics[width=0.8\textwidth]{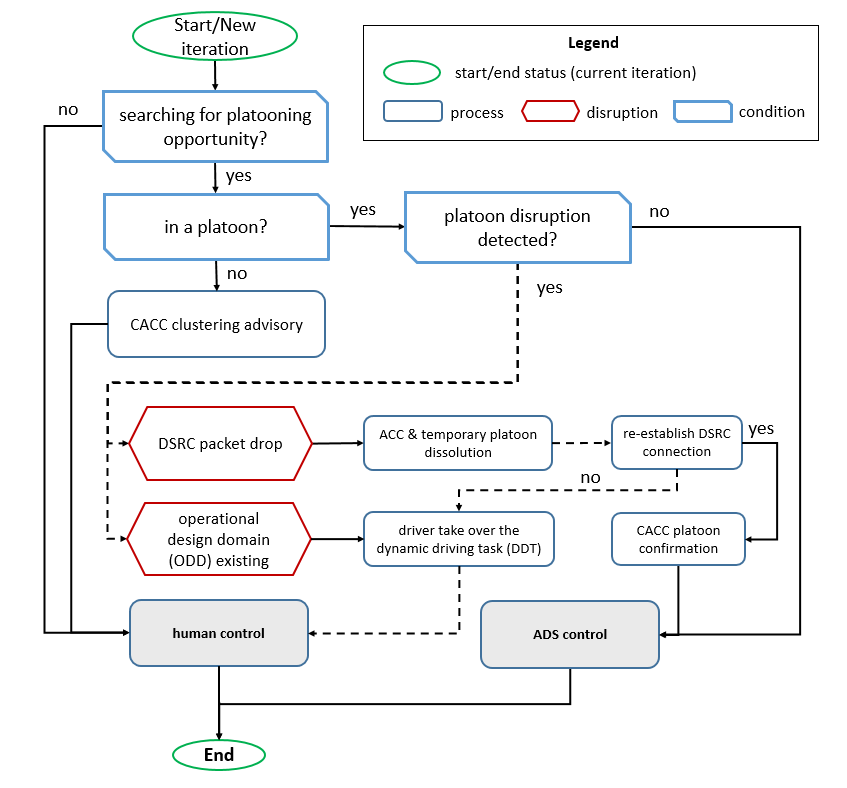}   
	\caption{\textcolor{revised}{CACC system fallback loop}}
	\label{fig:sysFallback}
\end{figure}

While the imperfect wireless communication is not the main focus in this paper, it is worth mentioning that an analytical probability module that was derived from a packet-level network simulator (i.e. ns-2 \citep{chen2007overhaul}) by \cite{killat2009empirical}  was also incorporated into the VEDM.  One-hop DSRC broadcast reception probability can be calculated during the simulation based on transmission power, transmission distance, and communication channel loading.  The outcomes of the communication have an influence on the CACC platoon operation. The details regarding the implementation of this module in VEDM is available in \cite{Zhong2018}.  The behaviors of a CACC vehicle in a platoon is dictated by the successful V2V communication. Here the CACC system is assumed with SAE Level 3 vehicle automation  that requires a receptive fallback-ready driver\citep{SAE2014}.
\textcolor{red}{
Fig. \ref{fig:sysFallback} illustrates the system fallback in the event of automated driving system (ADS) failure, such as loss of DSRC packets.  It is executed at each simulation time step on an individual vehicle basis, and it determines the transition of the control authority between the ADS and the human driver.}

\subsection{Managed Lane Scenarios}
Eligibility, accessibility, and pricing are the three primary factors in designing a ML \citep{mlPrimer}.  In this study, only the first two factors are considered.  A three-phase ML deployment for CACC is envisioned as covered in the literature review.  Subsequently, four ML strategies, shown in table \ref{table: MLs}, are formulated for testing. 
\begin{itemize}
\item{\textbf{Base case (BASE)}: This scenario serves as the base condition of the I-66 segment for this study. As stated previously, an HOV lane is implemented in the leftmost lane of the roadway and the network has been calibrated.}
\item{\textbf{Unmanaged lane (UML)}: In this strategy, the HOV lane was revoked, and current HOVs are treated as GP vehicles. CACC vehicles are not given priority use of any lane, and they operate along with general purpose (GP) vehicles/ HVs.}
\item{\textbf{Mixed managed lane (MML)}: This strategy aims to utilize the current lane use configuration of the network and provide priority usage of HOV lane for CACC vehicles. It investigates the effectiveness of introducing CACC vehicle to an existing ML facility.}
\item{\textbf{Dedicated CACC lane (DL)}:  The exclusive access to the leftmost lane for CACC vehicles is studied. A homogeneous CACC traffic is believed to be beneficial for the CACC operation. The merging impact of CACC vehicles toward the leftmost lane can be studied as well. Like in UML, the HOVs are treated as GP vehicles in this strategy.}
\item {\textbf{Dedicated CACC lane with access control (DLA)}: This strategy is essentially a dedicated CACC lane mentioned above but with access control where CACC vehicles are only able to move in or out of the ML at designated locations. Therefore, the weaving activities are aggregated at certain locations of the segment. It is formulated to insulate the CACC vehicle platoons form the potential impacts of the weaving activities. }
\end{itemize}

\begin{table}[!ht]
\centering
\caption{Scenario settings for two cases.} 
\begin{tabular}{ccccccc}
\hline
\backslashbox{Strategy}{Lane}&  4\textsuperscript{th}(leftmost) & 3\textsuperscript{rd} & 2\textsuperscript{nd} & 1\textsuperscript{st} & Access Control \\ \hline
BASE & H & G & G & G & N \\
UML & G+C & G+C & G+C & G+C & N\\
MML & C+H & G & G & G & N\\
DL & C & G & G & G & N\\
DLA & C & G & G & G & Y\\
\hline
\footnote{Footnote}{H-HOV, G-GP, C-CACC}
\end{tabular}
\label{table: MLs}
\end{table}

Since MLs are commonly designed to be left-lane concurrent with at-grade entry/exit, the weaving which requires crossing multiple GPLs could have a negative impact on mobility and safety of the overall traffic. Therefore, the weaving length of the incoming CACC vehicle to the leftmost managed lane has to be considered when it comes to a ML design. \cite{caltran2018HOV} suggested the minimal distance of 800 ft (243 m) per lane; whereas \cite{wsdot2006HOT} proposed the minimal distance of 500 ft (152m) between an access zone and immediate on- or off- ramp. The access control level is a trade-off among accessibility, efficiency, and safety. The optimal access control level was studied by \cite{saad2018access} and the weaving length of 1,000 ft (333m) per lane change was recommended. \cite{cai2018safety} investigated the safety impact of weaving lengths for lane change maneuver using a driving simulator. The 1,000 ft (333 m) optimal length for each lane change is based on the three evaluative measures: speed standard deviation, minimum time to collision, and potential collision events.  According to the aforementioned recommendations, 1,000-ft (333-m) weaving length per lane for the location of the ML entry is adopted in this study.  

\section{Simulation Result}
\label{Simulation Result}

The effectiveness of CACC-ML is discussed from the perspectives of 
overall network performance, safety, environmental impact, equity, and platooning performance.
%\begin{enumerate*}[label=\roman*)]
%\item overall network performance,
%\item safety,
%\item environmental impact,
%\item equity, and
%\item platooning performance.
%\end{enumerate*}
To factor in the variability of simulation, five replications are run for each combination of strategy and MP.

\subsection{Network Performance}
\begin{figure} [ht!]
	\centering
	\includegraphics[width=0.9\textwidth]{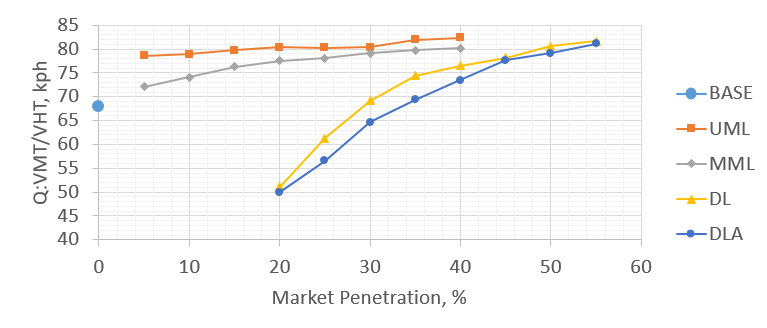}   
	\caption{Network Q value}
	\label{fig:qValue}
\end{figure}

The network-wide improvement of CACC under the four ML strategies are evaluated.  Being developed by Caltrans PeMS \citep{chen2001freeway}, the Q value is the ratio of vehicle mile traveled (VMT) and vehicle hour traveled (VHT). The VHT represents the input of a freeway system, whereas the VMT gauges the output of a freeway system. It can be also considered as the speed on an aggregated level. Hence, the higher the value of Q, the better performance or more productive of a highway system.  As shown in Fig. \ref{fig:qValue}, introducing CACC vehicles to the network helps to improve the network efficiency even when MP is below 15\%.  UML achieves the highest efficiency among all strategies across the MP range in terms of Q values. For low MP range (less than 30\%), the implementation of DL or DLA seems to be premature, because it decreases the efficiency of the network drastically. 
\begin{figure} [ht!]
	\centering
	\includegraphics[width=0.9\textwidth]{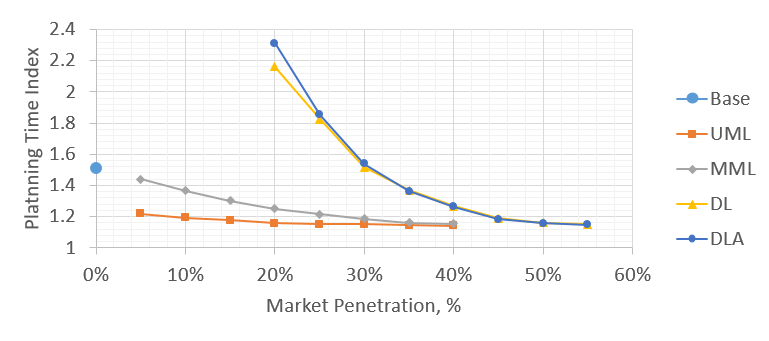}   
	\caption{Planning time index} 
	\label{fig:pti}
\end{figure}

The Planning Time Index (PTI) is defined as the ratio of the 95-percentile peak period travel time to the free-flow travel time \citep{reliability2006making}. It is used for evaluating the travel time reliability of the ML strategies. As shown in Fig. \ref{fig:pti},  the original PTI for traversing the network is 1.50, which indicate that a driver needs to allocate 1.5 times of the free-flow travel time in order to ensure a 95\% on-time arrival. The PTIs for UML and MML exhibit a decreasing trend as more CACC vehicles are introduced into the network.  At 40\% MP, the lowest PTIs for UML and MML are observed as 1.14 and 1.15, respectively.  Travelers in both DL and DLA experience higher PTIs than the BASE when MP is below 30\%. With a decreasing trend, at 50\% MP, the DL and DLA reach the same level of the PTIs as observed for UML and MML at 40\% MP.

\subsection{Safety}
\begin{figure} [H]
	\centering
	\includegraphics[width=0.9\textwidth]{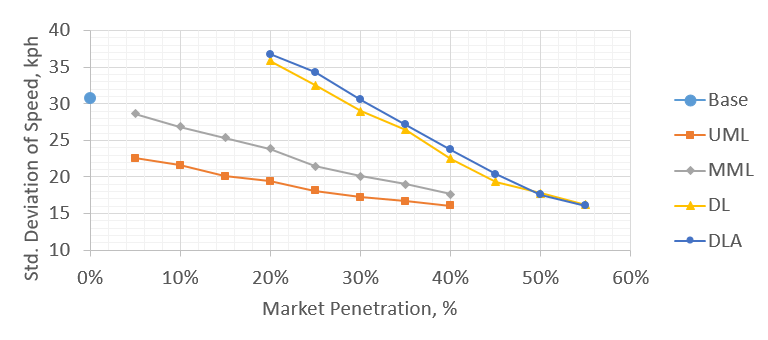}   
	\caption{Standard deviation of speed} 
	\label{fig:stdSpd}
\end{figure}

The standard deviation of speed, as a Safety Surrogate Assessment Model (SSAM) \citep{lee2014mobility, li2017evaluation}, has been used to assess the safety performance \citep{gettman2003surrogate}. Generally, a lower value in the standard deviation of speed represents a more stable traffic flow \citep{songchitruksa2016}. In any of the ML strategy, the standard deviation of speed exhibits a decreasing trend as the MP goes up as shown in Fig. \ref{fig:stdSpd}.  The improvement in terms of the SSAM is observed in UML and MML even at low MP range. The DL and DLA start with higher values of standard deviation than that of the BASE. They both resumes to the baseline level at 30\% MP. Between DL and DLA, the former display a marginally lower standard deviation at each MP level that is below 50\%. Additionally,  the speed standard deviation of UML is consistently lower than that of MML throughout all levels of MP.

\subsection{Environmental Impact}
\begin{figure} [h]
	\centering
	\includegraphics[width=0.9\textwidth]{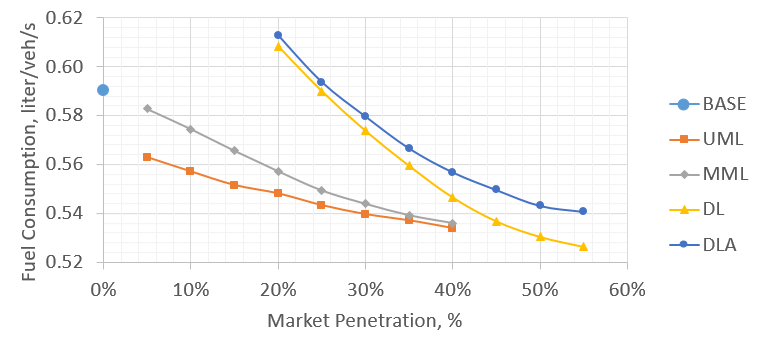}   
	\caption{Average fuel consumption} 
	\label{fig:fuelCon}
\end{figure}

The VT-Micro model \citep{rakha2004vtmicro} is implemented in the Vissim EmissionModel.DLL \citep{VissimEmitDll} to collect second-by-second fuel consumption for each vehicle. The entire network is treated as a system and the average fuel consumption for each vehicle per second is shown in Fig. \ref{fig:fuelCon}.  All the testing scenarios yield a lower fuel consumption with the exception of 20\% and 25\% MP for both DL and DLA.  The higher the MP, the greater the reduction in fuel consumption is observed. 

\subsection{Equity}
\begin{figure} [H]
	\centering
	\includegraphics[width=\textwidth]{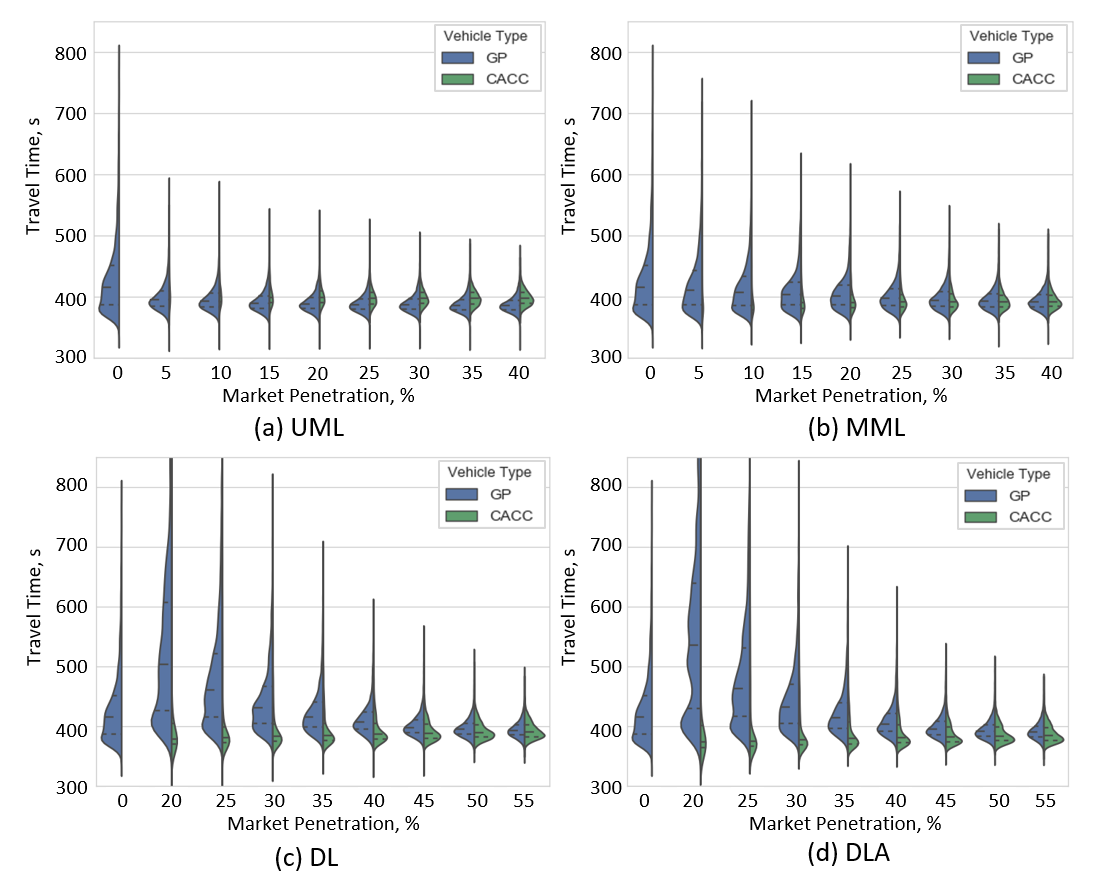}   
	\caption{Travel time comparison based on vehicle type} 
	\label{fig:equity}
\end{figure}

Equity refers to fairness when it comes to distributing impacts.  Horizontal transportation equity, in the context of CACC-ML, is the impact distribution among CACC users and non-CACC users. Traffic speed and roadway level-of-service can be used as indicators to measure impacts that affect equity \citep{litman2002equity}.  When the performance on GPLs is degraded, negative equity is distributed to GP users.  To measure the equity distribution among all four ML strategies, the segment travel time for traversing the entire segment on the mainline is compared. For each Violin plot shown in Fig. \ref{fig:equity}, the widths on either side are scaled by the number of observations. The quartiles of the distributions are also shown in the shade of the violin plot.  Under UML, a narrower distribution of travel time is observed even at 5\% MP, which means both CACC or GP vehicles do not experience negative equity.  Across the rest of the MP levels, the distributions of GP travel time remain nearly unchanged but with fewer high values observed in the BASE. With the increase of CACC vehicles, it is also observed that the median values of the distributions remain the same in UML. For MML where HOVs are grouped into GP vehicles, improvement only shows when MP reaches 10\% and the trend is maintained until MP reaches 50\%. For DL and DLA, the travel times get worsen when MP is below 30\% and the improvement only shows above 35\% MP.  Among all the scenarios, the CACC travel time distributions have lower variances and medians than GP vehicles that are operated in the prevailing traffic condition. 

\subsection{Platooning Performance}
\begin{figure} [!ht]
	\centering
	\includegraphics[width=0.9\textwidth]{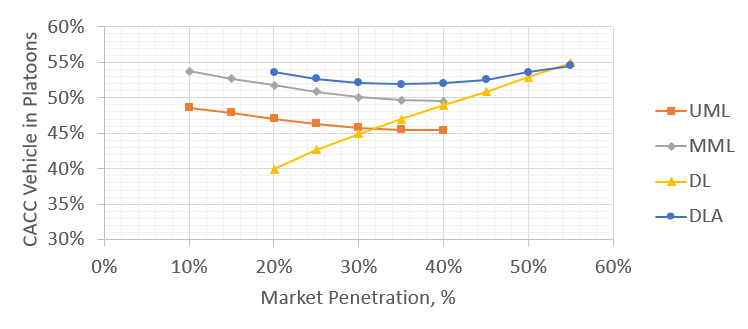}   
	\caption{Percentage of CACC vehicle in platoons} 
	\label{fig:pltnPrec}
\end{figure}

In this section, a set of rarely-examined MOEs is introduced. They are related to CACC clustering and made possible by the high customizability of VEDM. One of the major benefits of CACC is the compact vehicle platoons with short intra-platoon following distance enabled by the V2V communication. As such, the number of vehicles within platoons is positively correlated to the improvement in throughput. The average percentage of CACC vehicles is shown in Fig. \ref{fig:pltnPrec}. With the CACC demand increases, the percentage of the number of platooned CACC vehicles decreases in both UML and MML, which seems counterintuitive. In fact, this indicates the increased difficulties for additional CACC vehicles to form platoons. When the slope of a curve is positive as shown in DL, the platooning opportunity increases as the presence of CACC vehicles increases. If a curve linearly increases at a 45-degree slope, it is an indication that all the additional CACC vehicles get the same likelihood of forming platoons.  For a negative slope of the curve, the corresponding ML strategy is not able to offer clustering opportunity at the same level as MP grows. Between 20\% and 50\% MP, DLA outperforms other strategies and it is only marginally surpassed by DL at 55 \% MP.

\begin{figure} [!ht]
	\centering
	\includegraphics[width=0.9\textwidth]{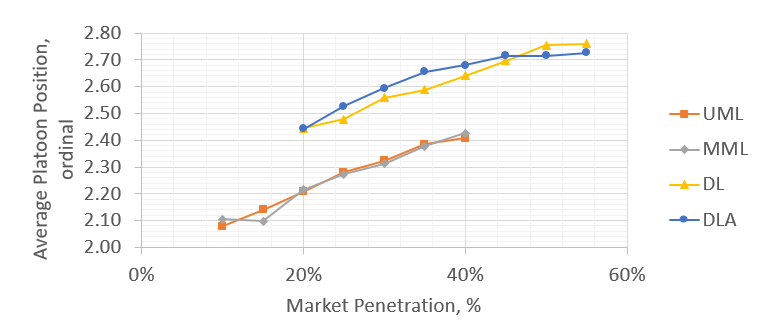}   
	\caption{Average platoon position} 
	\label{fig:avgPltnPos}
\end{figure}

During the simulation run, each CACC vehicle registers its sequence in a platoon.  Due to the decentralized execution for each CACC vehicle in VEDM, the mean platoon depth \citep{Amoozadeh2015} is calculated. The mean platoon depth represents the mean platoon position reported by each CACC vehicle and it is a good indicator of the average size of platoons in the network.  As shown in Fig. \ref{fig:avgPltnPos}, there is a major distinction between the two types of ML strategies: the type with dedicated CACC lane (i.e. DL and DLA) and the type without. The smallest and greatest gaps observed between these two types of ML strategies are 0.16 at 25\% MP and 0.25 at 35\%, respectively.  

\begin{figure} [!h]
	\centering
	\includegraphics[width=0.9\textwidth]{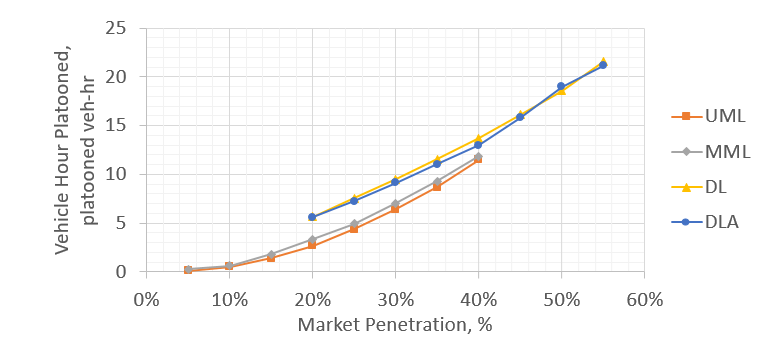}   
	\caption{Vehicle hour platooned} 
	\label{fig:vhp}
\end{figure}

The formation and dissolution of CACC platoons occur dynamically under mixed traffic conditions due to various reasons (e.g. vehicles exiting a freeway, non-CACC vehicles cutting in, or communication disruption).   At certain timeframe, there is a percentage of CACC vehicles that are in platoons, utilizing the intra-platoon, short following distance. On the other hand, there is the duration of the CACC vehicles in platoons. The dynamic changing of the two aforementioned measures during operation makes it difficult to spatially and temporally measure the effectiveness of forming CACC platoons.  To account for this phenomenon, the vehicle hour platooned (VHP) is proposed in this study to quantify the temporospatial platooning performance for the entire network. The VHP is the accumulation of the number of CACC vehicles that are in platoons within an hour. In Fig. \ref{fig:vhp}, the convex shape in UML and MML curves indicate that a non-linear growth of VHP as the MP increases.  Compared to UML and MML, DL and DLA yield consistently higher values of VHP. The increasing trends in DL and DLA exhibit a linear pattern.

A score matrix system that factors in all the aforementioned performance measures is developed. Table \ref{table: scoreMat} summarizes the score assignment rule. The first four heterogeneous types of performance measures (i.e. mobility, safety, equity, and environmental impact) are determined based on the comparison with the BASE, whereas the CACC platooning score is determined by the comparisons among the four tested strategies at each level of MP. 

\begin{table}[!ht]
\centering
\caption{Vehicle Control Parameters} 
\begin{tabular}{c|c}
\hline
Parameter & Evaluation Score \\ \hline
Traditional performance measure & improvement: 1 \\
(mobility, safety, equity, & neutral: 0 \\
and environmental impact) & degradation:-1 \\
  \\ \hline
 & ranked among 4 strategies  \\ 
CACC platoon formation & 1st: 4, 2nd: 3, 3rd: 2, 4th: 1  \\
\hline

\end{tabular}
\label{table: scoreMat}
\end{table}

\begin{figure}[h!]
\centering
\subfloat[traffic performance score]{\label{main:a}\includegraphics[scale=.5]{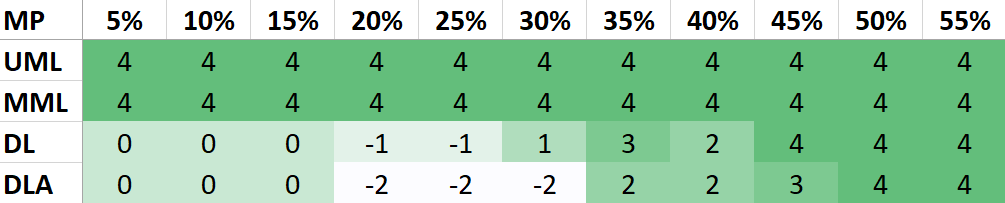}} \par\medskip

\begin{minipage}{.45\linewidth}
\centering 
\subfloat[platooning performance score]{\label{mainb}\includegraphics[scale=.45]{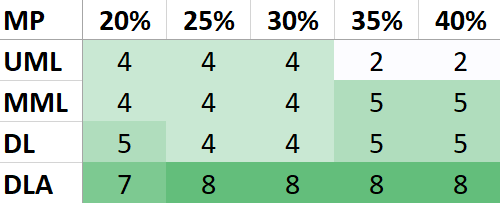}}
\end{minipage} 
\begin{minipage}{.45\linewidth}
\centering
\subfloat[normalized sum of score]{\label{main:c}\includegraphics[scale=.45]{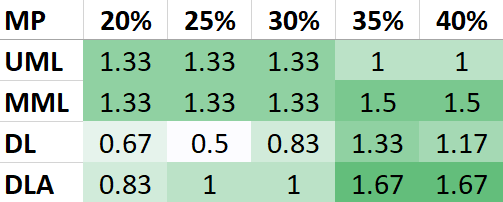}}
\end{minipage}  
\caption{Managed lane evaluation score}
\label{fig:mlScore}
\end{figure}

\begin{figure} [!htb]
	\centering
	\includegraphics[width=0.8\textwidth]{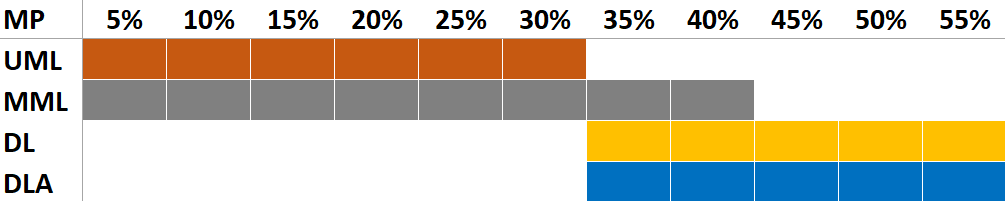}   
	\caption{Managed lane policy recommendations} 
	\label{fig:mlRecom}
\end{figure}

Fig. \ref{fig:mlScore}(a) and (b) show the evaluation scores for the network performance and platooning performance, respectively. Since the calculation of the two score categories is different, normalization is performed before summation to obtain the scores in Fig. \ref{fig:mlScore}(c) that  represent the overall evaluation scores between the critical MP range. When MP is below 25\%, both UML and MML strategies are suitable options for CACC deployment, depending on whether an existing ML strategy is already in place.  When the MR reaches above 30\%, both MML and DLA strategies are viable options. Either DL or DLA is able to create homogeneous CACC traffic; however, DLA is able to provide better operational accommodation for CACC. The recommendations are summarized in Fig. \ref{fig:mlRecom}.

%\begin{figure}
%\begin{subfigure}{0.4\linewidth}
%\centering
%\includegraphics[keepaspectratio=true]{figure/scoreA.png}
%\caption{}
%\label{fig:sub1}
%\end{subfigure}%
%\begin{subfigure}{0.4\linewidth}
%\centering
%\includegraphics{figure/scoreB.png}
%\caption{}
%\label{fig:sub2}
%\end{subfigure}\\[1ex]
%\begin{subfigure}{0.4\linewidth}
%\centering
%\includegraphics{figure/scoreC.png}
%\caption{}
%\label{fig:sub3}
%\end{subfigure}
%\caption{Three subfigures}
%\label{fig:test}
%\end{figure}

\section{Conclusions}
\label{Conclusions}
This study focuses on assessing the effectiveness of four managed lane (ML) strategies for the near-term deployment of CACC under mixed traffic conditions.  An integrated microscopic simulation testbed is employed to conduct the large-scale traffic simulation on the calibrated I-66 segment in Fairfax County, Virginia.  Note that the types of the ML strategies tested in this study are not intended to be exhaustive. 
\textcolor{red}{
Additionally, the control parameters for CACC vehicles could vary when it comes to implementation. The desired time gaps used in this study are relatively conservative (i.e., 1 s and 1.2 s), and it would become even shorter as the MP grows. The powertrain parameters used in this study could also be different in the future as lighter vehicles (without the heavy safety frame) become viable due to drastically fewer crashes. 
}
Nevertheless, the simulations reveal several important aspects of deploying CACC in mixed traffic, most of which have rarely been discussed and yet the stakeholders have to take into account. Besides the traditional network-wide performance measures, a set of platoon performance measures are put forward to evaluate the platoon formation of CACC vehicles.  Additionally, the evaluation score matrices considering various aspects of CACC deployment were proposed to demonstrate the overall suitability for each ML strategy. 

For locations without existing managed lane policy, mixing CACC traffic across all travel lanes is an acceptable option when the market penetration rate is below 30\%. Mixed managed lane (e.g., HOV + CACC, HOT + CACC) is a versatile option for providing priority lane usage for CACC vehicles, especially in the locations where ML strategies have already been implemented. It is premature to implement CACC lane when market penetration (MP) is below 30\%, due to the degradation of the traffic condition on the general purpose lanes. When the mid-range MP (30\% to 55\%) is reached, CACC lane begins to show its advantages for facilitating CACC clustering and forming homogeneous CACC traffic that is ultimately necessary for the further harness of the benefits of CACC (e.g., higher speed limit and short headway intra-platoon car following).

Further exploration of the operation of CACC-ML is very much needed, including the ways to optimize the operational efficiency for the managed lane in conjunction with other CAV-based systems, such as speed harmonization, automated merging assistance, one-way vehicle communication devices, etc. In addition, the clustering strategy for CACC-ML in mixed traffic conditions is an under-explored area at the current stage. Such operation should be performed at a tactical driving level beyond the basic operational level as defined by \cite{SAE2014}. Moreover, the interactions between CAVs and HVs is another area that deserve more attentions. Thus far, the majority of the studies have been focusing on CAVs. The potential impacts on HVs from CAVs (e.g., platoon blockage, induced weaving) should not be overlooked, especially when active clustering strategy is employed.

\newpage
\appendix
\section{}
Definitions for the abbreviations used \\ \newline
%\begin{table}[H]
%\centering
%\caption{\textcolor{revised}{List of Abbreviations}} 
%\resizebox{0.9\textwidth}{!}
%{
\begin{tabular}{p{2in}|p{4in}} 
\hline  
\textbf{Abbreviation} & \textbf{Definition} \\ \hline 
V2V & vehicle-to-vehicle\\ \hline 
DOT & department of transportation \\ \hline 
CACC & cooperative adaptive cruise control \\ \hline 
CAV & connected and automated vehicle\\ \hline
DSRC & dedicated short-range communication\\ \hline
MP & market penetration \\ \hline
HOV & high occupancy vehicle\\ \hline
HOT & high-occupancy toll \\ \hline
XBL & exclusive bus lane \\ \hline
GP & general purpose vehicle\\ \hline
HV & human-driven vehicle\\ \hline
AHS & automated highway system\\ \hline
ADS & automated driving system \\ \hline
SARTRE & Safe Road Trains for the Environment project\\ \hline
PATH &  Partners for Advanced Transportation Technology\\ \hline
ML & managed lane\\ \hline
RFID & radio frequency identification   \\ \hline
GPS & global positioning system \\  \hline
COM & component object model    \\ \hline
VEDM & Vissim external driver model    \\ \hline
E-IDM & enhanced intelligent driver model    \\ \hline      
CAH & constant-acceleration heuristic \\ \hline 
MOBIL & minimizing overall braking induced by lane change\\ \hline 
UML & unmanaged lane \\ \hline 
MML & mixed managed lane \\ \hline 
DL & dedicated lane\\ \hline 
DAL & dedicated lane with access control\\ \hline 
VMT & vehicle mile traveled \\ \hline 
VHT & vehicle hour traveled \\ \hline 
PTI & planning time index \\ \hline 
SSAM & safety surrogate assessment model \\ \hline 
MOE & measure of effectiveness \\ \hline 
VHP & vehicle hour platooned \\ \hline 
DDT & dynamic driving task \\ \hline
ODD & operational design domain  \\ 
\hline    
\end{tabular}
%}
%\label{table:abbrv}
%\end{table}

\bibliographystyle{elsarticle-harv}
\bibliography{CACC-ML_TRA_2019_323_R2}
\end{document}